\documentclass[usenatbib]{mn2e}
\usepackage{graphicx}

\begin{document}

\newcommand{\mbh}{M_{\bullet}}
\newcommand{\msun}{M_\odot}

\title[Accretion rings revealed by future X-ray spectroscopy]
{Black hole accretion rings revealed by future X-ray spectroscopy}

\author[V. Sochora, V.~Karas, J. Svoboda, and M.~Dov\v{c}iak]
{V. Sochora, V. Karas, J. Svoboda, and M. Dov\v{c}iak
 \\~\\
 Astronomical Institute, Academy of Sciences of the Czech Republic,
 Bo\v{c}n\'{\i}~II 1401, CZ-141\,31~Prague, Czech Republic}

\date{Accepted 2011 July 20; Received 2011 July 20; in original form 2011 January 25}
\pagerange{\pageref{firstpage}--\pageref{lastpage}}
\pubyear{2011}
\maketitle
\label{firstpage}

\begin{abstract}
Spectral features can arise by reflection of coronal X-rays on a black hole accretion 
disc. The resulting profile bears various imprints of strong gravitational field acting 
on the light emitting gas. The observed shape of the reflection line is formed by 
integrating contributions over a range of radii across the accretion disc plane, where the 
individual photons experience different level of energy shifts, boosting, and amplification 
by relativistic effects. These have to be convolved
with the intrinsic emissivity of the line, which is a function
of radius and the emission angle in the local frame. 
We study if the currently discussed instruments on-board X-ray
satellites will be able to reveal the departure of the line radial emissivity 
from a simple smooth power-law function, which is often assumed in data
fitting and interpretation. Such a departure can be a result of excess
emission occurring at a certain distance. This could be used to study
variations with radius of the line production or to constrain the position of 
the inner edge of the accretion disc. By simulating artificial data 
from a bright active galactic nucleus of a type-1 Seyfert galaxy 
(inclination $\simeq30$~deg, X-ray flux $\simeq1$--2 mCrab in keV energy 
band) we show that the required sensitivity and energy resolution could
be reached with Large Area Detector of the proposed LOFT mission.
Galactic black holes will provide another category of potentially suitable 
targets if the relativistic spectral features are indeed produced by reflection
from their accretion discs.
\end{abstract}

\section{Introduction}
Various pieces of
evidence support the idea that the X-ray emission from active galactic
nuclei \citep[AGN; see][]{fabian00,reynolds03} and stellar-mass black-holes
\citep{miller02,mcclintock06} originates from an accretion disc and the
surrounding corona. These are thought to be situated near a central
black hole, no more than a few tens gravitational radii from the event
horizon, giving rise to the relativistic effects
in 6--7 keV iron line complex and the underlying continuum. The standard
scheme of accretion discs \citep{novikov73,page74} captures the
main properties of accreting black holes surprisingly
well, nevertheless, the model omits some important aspects. In
particular, the radial profile of the source intrinsic emissivity is
represented by a smooth function (decaying as a power law $\propto
r^{-3}$ at large distance), while the realistic profile is likely to be
more complicated.

Different approaches have been pursued in order to understand how 
accretion disc X-ray spectra are formed. Generally, these include 
the investigations of accretion disc instabilities as well as the 
interpretation of spectra
to constrain the model parameters (such as the black hole spin, the
source orientation, and the location and size of the accretion disc), 
and to determine the radial profile of accretion disc emissivity
\citep{fabian04}. Recently, \citet{wilkins11} discussed an 
interesting approach to the inversion problem of determining the 
radial emissivity bulk profile of the relativistic broad iron line in 
Seyfert galaxy 1H 0707-495.
Here we turn our attention to additional features superposed
on top of the broad line.

The form of the inner accretion flow remains an open question. 
According to the standard scheme the flow proceeds down to the inner
edge at the innermost stable circular orbit (ISCO), i.e., $r_{\rm
ISCO}=6\,r_{\rm g}$ for Schwarzschild black hole, and $r_{\rm
ISCO}=1\,r_{\rm g}$ for an extreme Kerr black 
hole (gravitational radius
$r_{\rm{}g}\,\equiv\,GM/c^2\,\doteq\,1.48\times10^{12}M_7~\mbox{cm}$
with $M_7$ being the mass of the supermassive black hole in units of
$10^7M_{\odot}$). There are uncertainties about this
assumption. It has been argued that because of magnetic stresses 
the radiation edge of the emission may be
somewhat off the ISCO, and its exact location may be different for the
continuum and for the iron line \citep{reynolds08,abramowicz10}. Also, the 
inner rim recedes further out from the
black hole when the accretion flow ceases and the disc becomes 
truncated, as was demonstrated in several objects 
\citep[][and references cited therein]{markowitz09,svoboda10}.

Intermittent episodes of a localized disc irradiation
can naturally lead to a radially
stratified emission profile rather than monotonic, 
continuous dependence of a standard accretion disc. 
We can approximate this configuration by radially constrained 
zones, which can be called ``rings''. Let us remark that 
we are concerned with a spectral 
line emissivity, which is only partly related to the gas density;
the essential quantity here is the ionization state of matter
and how this varies with radius. Localized coronal irradiation of
the disc material enhances the line emission above the mean 
value in the neighbourhood of a certain point. Integration of detected signal
over a period of time then effectively produces a ring-type source 
\citep{goosmann07}.

In other words, one need not
imagine physically separate rings and gaps emerging
within the disc, even if this possibility has been also discussed; e.g.\
\citet{cuadra09} show a temporary density ring in an accretion
disc surrounding a black-hole binary, while \citet{artymowicz93} and
\citet{karas01a} examine the process of gap formation by a secondary 
satellite embedded within the disc.

The excess emission from a certain radius could be
revealed by high-resolution spectroscopy of the broad iron line
\citep{karas10}, although at present it appears to be
rather unfeasible because of insufficient
signal-to-noise in the available data. In this paper we discuss
whether the proposed Large Observatory for X-ray Timing 
(LOFT; \citeauthor{feroci10} \citeyear{feroci10}, \citeyear{feroci11}) will
have the necessary capability, at least for bright enough sources. 
To this end we produce artificial
data with appropriate properties and then we analyze 
them by using a preliminary response file.

\section{Reconstructing parameters from model spectrum}
\subsection{Local peaks of the broad line profile}
According to the standard (stationary) accretion disc scenario, most 
radiation is produced within just a few gravitational radii above ISCO 
(up to $\simeq20r_{\rm g}$). 
Also various X-ray spectral features and the power-law component
of the disc corona are believed to originate in
that region. This main flux-producing area of the accretion
disc is still rather wide in radius, hence it is thought to be the origin 
of the broad component of the line, which we include by employing the 
relativistic {\tt kyrline} model \citep{dovciak04}.

Some reports suggest transient structures that are localized in 
radius and exhibit themselves as narrow features in the spectrum
\citep[e.g.][and references cited therein]{guainazzi03,dovciak04a,demarco09}.
In our picture of time-integrated spectra these are represented by rings. 
Although the significance of these features is still
debated, their origin fits in the scenario of magnetic flares
as sites of primary local illumination.

Magnetic flares are thought to occur above the accretion disc 
\citep{czerny04,uzdensky08}; in this context the line emission of the
rings would arise not because the accretion disc develops such a physical
structure, but instead the localized emission by reconnection
flares illuminates the underlying disc at a certain place, affecting there its
ionization and producing the observable reflection features in spectrum. 
Rotation of the disc matter and time integration
of the signal during observation produce belts out of point-like
sites of illumination. In fact, we could then call these structures
as reflection rings.

Although the mentioned scheme is our preferred way of effectively
producing the rings, similar effects 
can arise by different mechanisms operating near a black
hole. They can be broadly distinguished in two categories. Firstly,
during the period of intermittent accretion the flow of material
varies and the density profile and other variable (ionization state profile) 
can change along \citep[e.g.][]{king06,ballantyne11}.
In another context, global spiral waves were suggested as a possible source
of light curve modulation of accreting black holes \citep{tagger06}.
Unlike ideal rings these spirals structures have a non-neglibible
radial extent, although to certain extent they could mimic rings in 
case of significant wounding in azimuthal direction \citep{karas01}.
The formation of separate annuli occurs also in some
models of strongly magnetized plasma discs \citep{coppi06}. 
Finally, confined radii of enhance emission are a consequence
of avalanche models (Pech\'a\v{c}ek et al., in preparation) 
where multiple flares arise in mutually connected
families and gradually propagate towards the center before 
decaying at some characteristic distance.

In contrast to the model
of major flares mentioned formerly, the latter alternatives
appear to be unable of modulating the outgoing spectrum
in any significant manner for a required period of time 
\citep{schnittman06}. Nonetheless, the current 
modelling has not yet provided a definitive answer and different
options should be investigated further. In fact, the topic for debates
is the expected level of the electromagnetic signal modulation above
the noise and
the ability to detect it by a given technology, rather than the existence
of the modulation at a certain small level.

Aside from the mentioned speculations, one may ask if the actual
presence of the rings can be tested. Here we thus
presume that the structure develops and modulates the outgoing
signal, and we want to verify whether the location of emission rings
could be reconstructed from the observed spectrum.
By {\em assuming the existence} of the modulation we
investigate if these structures can be spectrally 
resolved in present-day or future X-ray spectra.

\begin{figure*}
\begin{center}
\includegraphics[angle=-90,width=.49\textwidth]{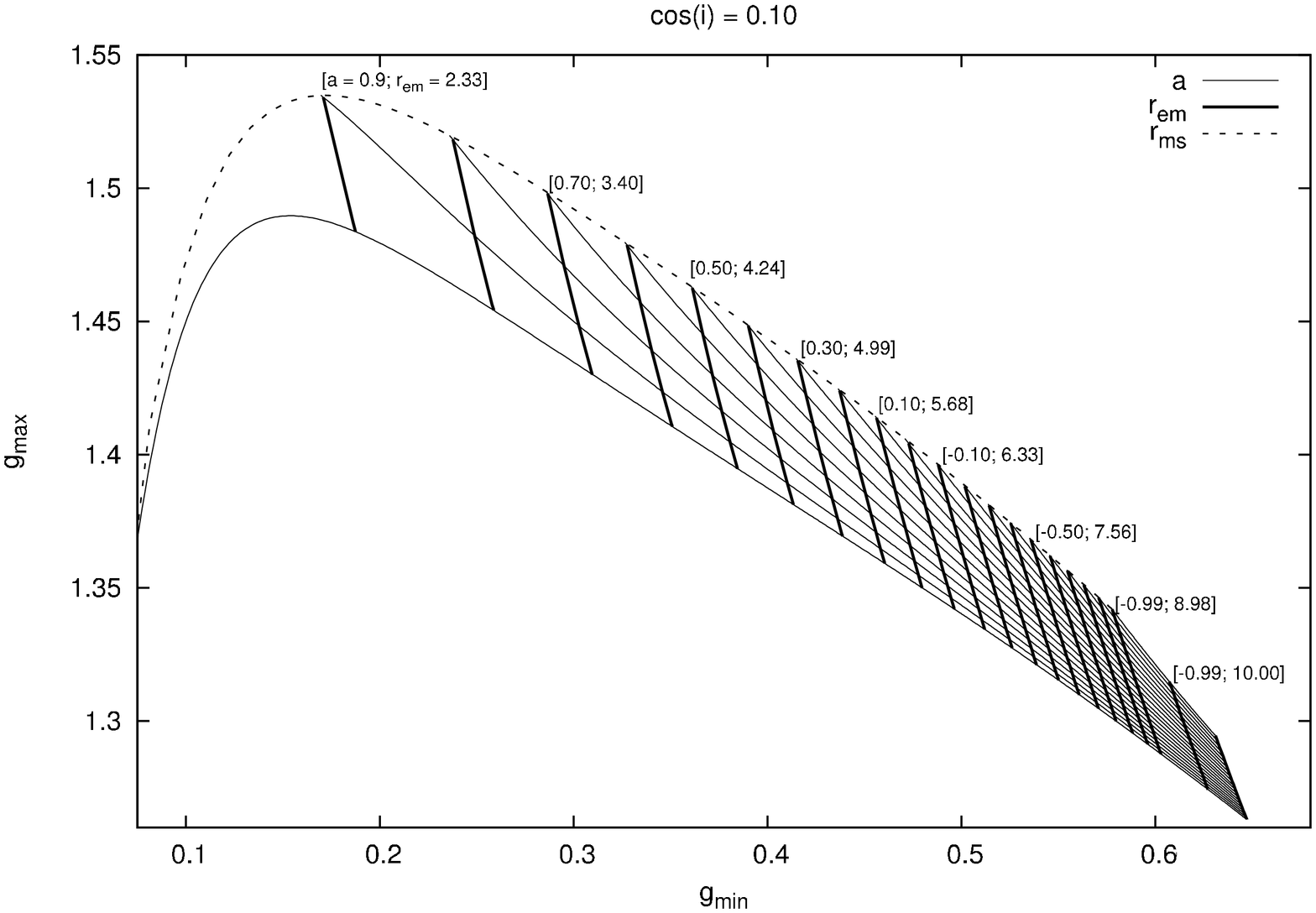}
\hfill
\includegraphics[angle=-90,width=.49\textwidth]{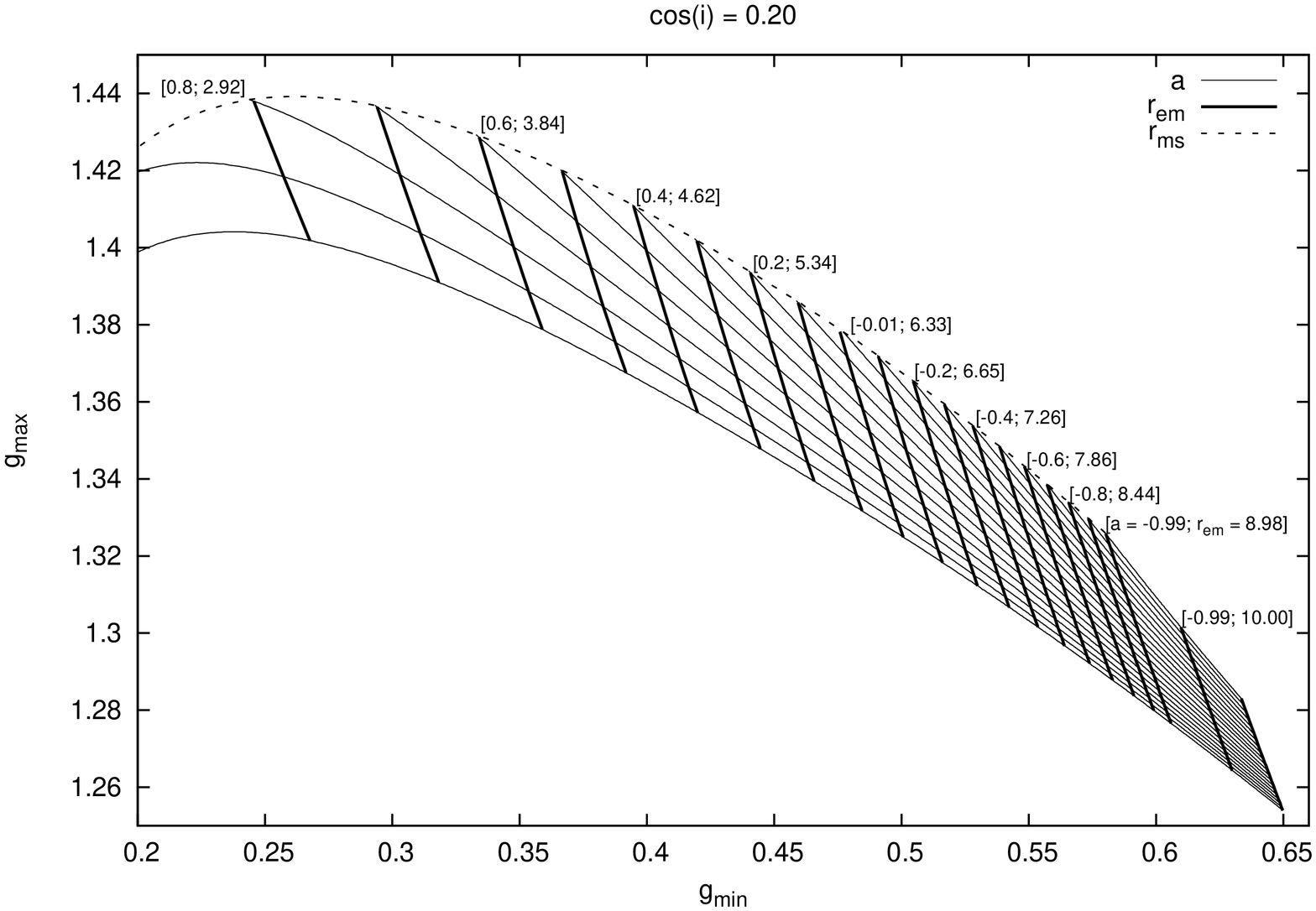}\\~\\
\includegraphics[angle=-90,width=.49\textwidth]{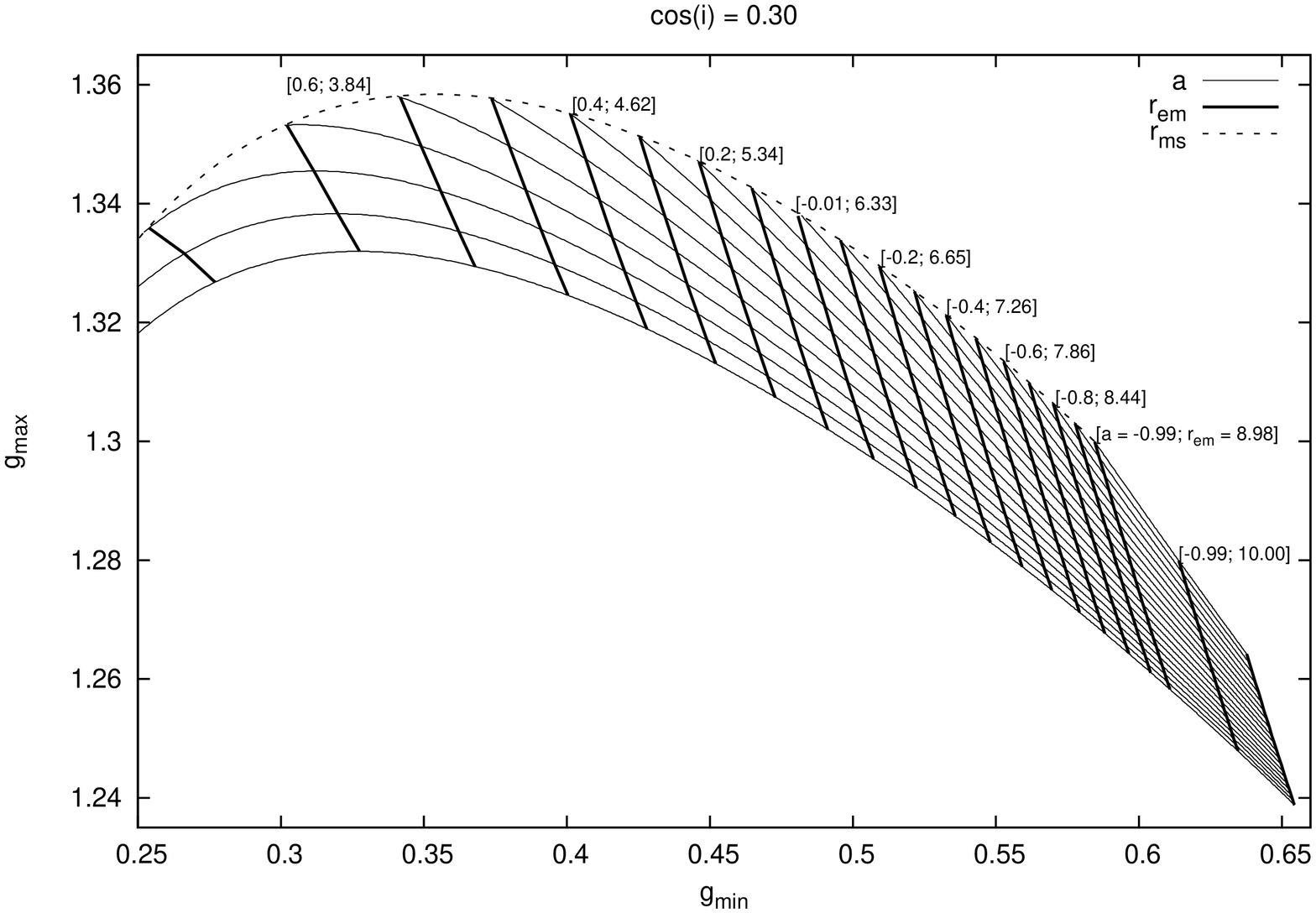}
\hfill
\includegraphics[angle=-90,width=.49\textwidth]{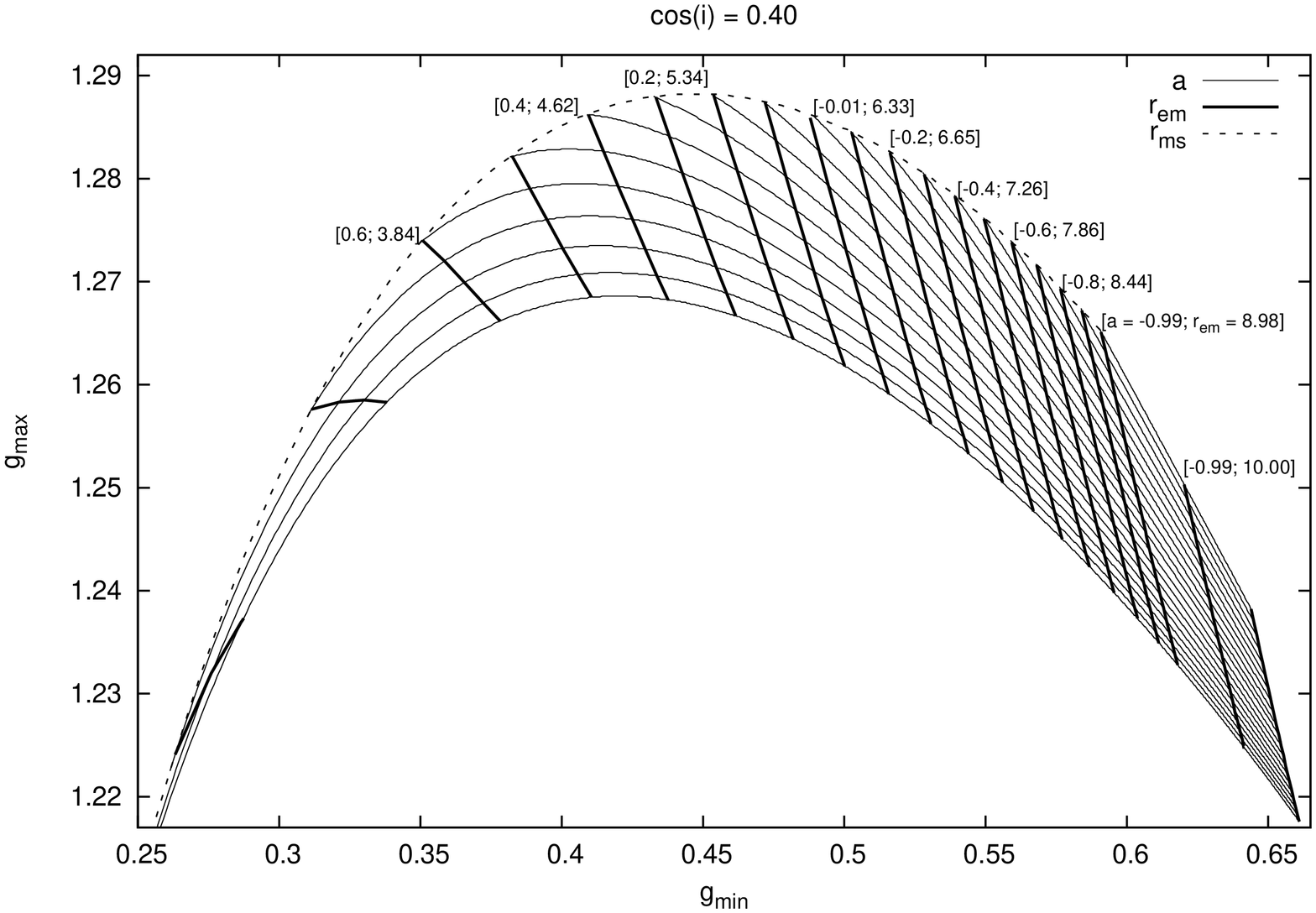}
\caption{Observed energy of the line peaks, as determined by the extremal 
shifts $g_{\rm max}$ versus $g_{\rm min}$
for different view angles of the observer, $\cos i$. Each pair of
$g_{\rm{}max}$, $g_{\rm{}min}$ values corresponds to a certain emission
radius, $r=r_{\rm{}em}$, and the black hole spin $a$. Curves of constant
emission radius  $r_{\rm{}em}$ and the black hole spin $a$ are shown
(values are given in brackets). The ISCO radius coincides with the
marginally stable orbit, $r=r_{\rm{}ms}$, and it defines
one boundary of the plot (dotted curve).
\label{fig:g}}
\end{center}
\end{figure*}

\subsection{The method of calculation}
To determine the relativistic shift (Doppler and gravitational) of the photon 
energy, we adopt the method described in our recent paper 
\citep{karas10}, where we considered propagation of photons from the
source in the limit of geometrical optics in Kerr metric
\citep{kato98}. The extremal values of energy shifts, $g_{\rm{}max}$ and
$g_{\rm{}min}$, are of particular interest, as they determine the range
of line energy observed from distance. The main parameters
are the emission radius
$r=r_{\rm{}em}\geq r_{\rm{}ISCO}$, the dimensionless spin parameter $a$
($-1\leq a\leq1$), and the inclination angle $i$ ($i=0$ stands for the
rotation axis, while $i=90^{\rm{}o}$ is for the edge-on view of the
accretion disc). Extending our previous work, we allow also the 
case of retrograde rotation ($a<0$)
because such a possibility is not excluded a priori, and it may lead
naturally to rings that emerge at somewhat larger radii 
than in the prograde case \citep[the ISCO
recedes as counter-rotation of the black hole accelerates toward 
$a=-1$;][]{dauser10}.\footnote{In the case of accretion discs 
counter-rotating with the black hole, we found a disagreement 
between our energy shifts compared with the corresponding 
values in \citet{dauser10}. We checked that the discrepancy 
up to $\sim3$\% (gradually growing with the spin and inclination angle) 
can be attributed to the way how frame-dragging effects had been 
treated in the original version of the {\tt relline} code. It turns out that 
this difference has been meanwhile eliminated from the latest version 
of the code, so the results are now in full agreement with the 
{\tt kyrline} \citep{dovciak04} version for retrograde spin. Naturally, 
in the latter case, ISCO can recede 
as far as $r=9GM/c^2$, and so the relativistic effects become quite 
small compared to discs rotating in the prograde direction.}

A simple prescription for the intrinsic emission of light allows us to
explore the observed spectral features across the parameter space. We
assume a power-law continuum (representing the primary X-rays
originating from corona) plus a spectral line
(representing the K$\alpha$ emission line of iron at $6.4$~keV rest
energy). The line is intrinsically narrow (in the
local co-moving frame) of the line-producing ring, although 
it becomes subsequently broadened by relativistic effects. 
The line is produced in a range of radii over the inner disc,
so the rings are just those radii where the line production is
enhanced above the baseline model of the power-law component and the 
broad line. The mentioned components of the
model spectrum give us an opportunity to test the procedure of
reconstructing the source emissivity. Light rays propagate
along null geodesics in the curved spacetime, which brings significant
energy shifts to the final spectrum and it spreads the observed profile
over the whole $\langle g_{\rm{}min},g_{\rm{}max}\rangle$ interval with
respect to the rest energy of the line. 

Figure \ref{fig:g} presents the extremal energy shifts, $g_{\rm{}max}$ vs.\
$g_{\rm{}min}$. These determine the expected energy of the local peaks
of the composite broad line, 
depending on the radial profile of emissivity from the disc.
One can see a complex interplay between gravitational effects,
Doppler shift and light-bending. The overall gravitational redshift
dominates near the horizon (although we restrict the emission radii to
$r>r_{\rm{}ISCO}$), while Doppler broadens the line (i.e.\ it
increases $g_{\rm{}max}/g_{\rm{}min}$ ratio) and it enhances the height
of blue wing, progressively more for large inclinations. The focusing
effect further enhances the observed flux, especially at large~$i$.

\begin{figure*}
\begin{center}
\includegraphics[angle=0,width=.48\textwidth]{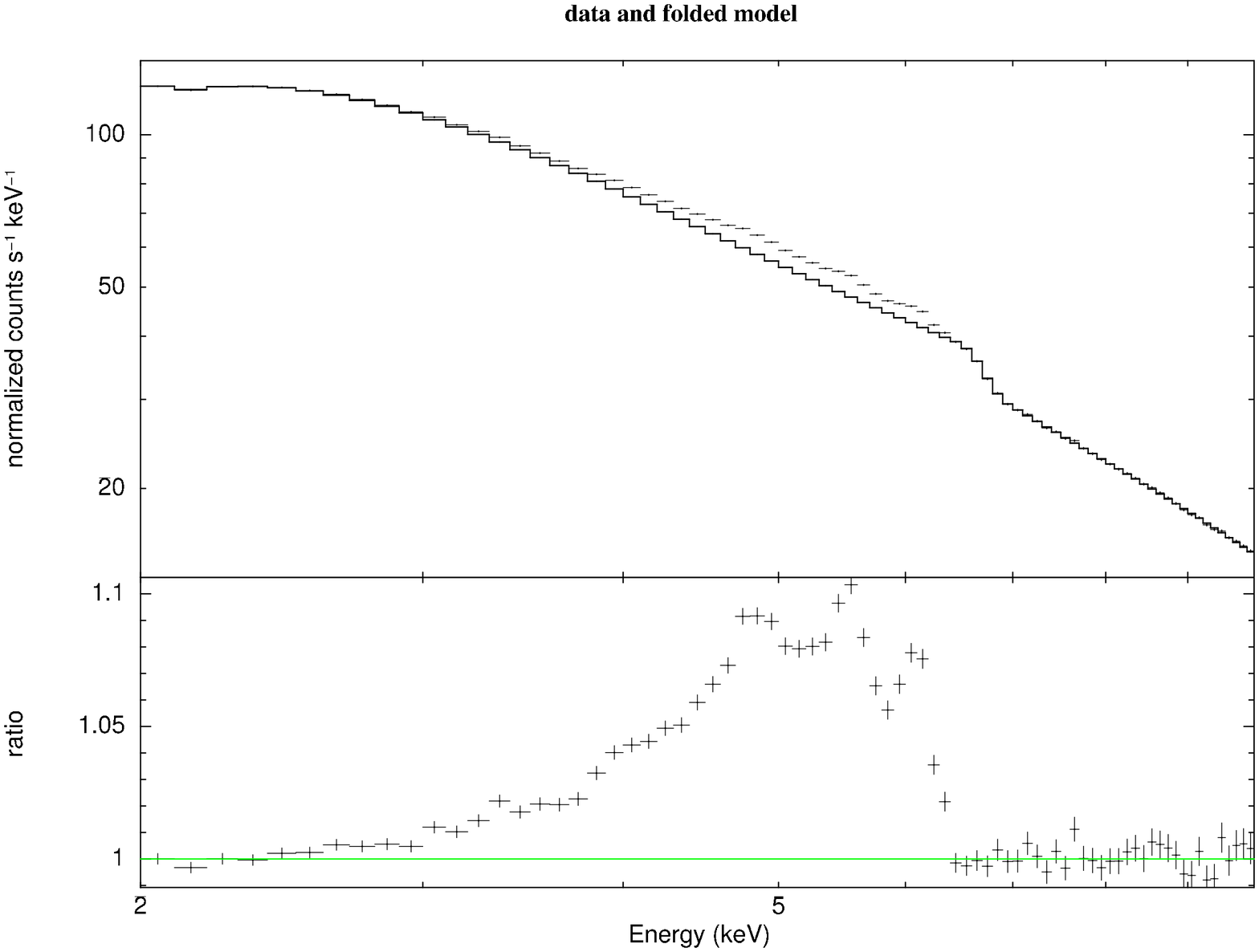}
\hfill
\includegraphics[angle=0,width=.47\textwidth]{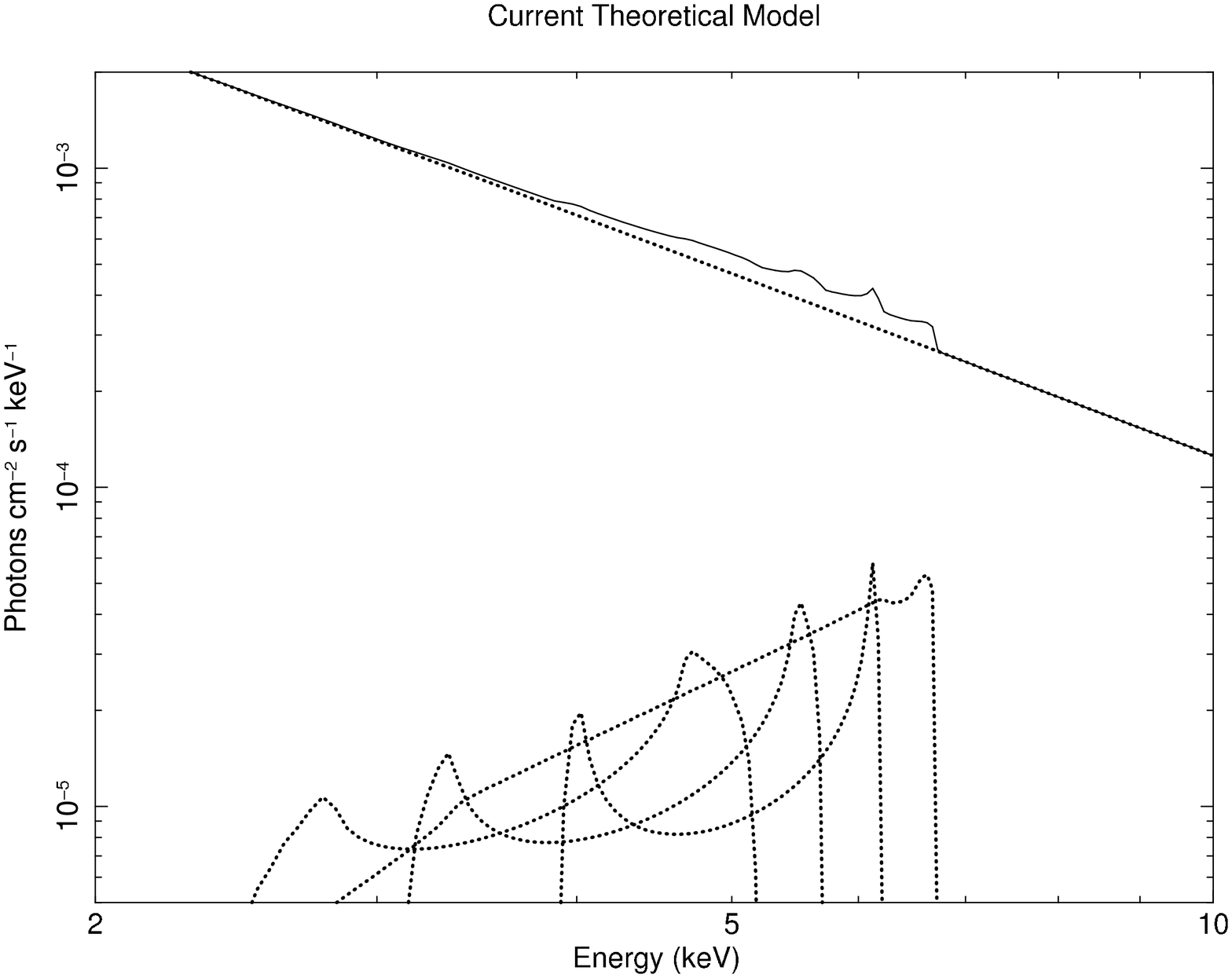}
\caption{Left panel: Simulated data and the ratio
to the the baseline model consisting of the power-law and the disc-line components
(background subtracted).
Residuals related to the three additional narrow rings are clearly visible.
Right panel: The complete theoretical model and the model components:
a power-law continuum and the individual line profiles from which
the energy shifts of the components are derived.
\label{fig:ringspectrum}}
\end{center}
\end{figure*}

Fig.~\ref{fig:g} provides several examples including both prograde 
and retrograde rotation, although we cannot show here all 
combinations of the parameters. More examples of the prograde
case were examined in \citet{karas10}; in particular; in that work
we show also the case of small inclination angles, which are 
relevant for type-1 AGNs but are more difficult to plot because the
energy shifts are much smaller too. 
From the figure one can also realize that the parameter degeneracy
of the model can be expected where the lines converge to each 
other, quite independent of the parameter values (we will come back
to this issue in more detail later in sec.~\ref{sec:Discussion}).

\begin{figure}
\begin{center}
\includegraphics[angle=0,width=.48\textwidth]{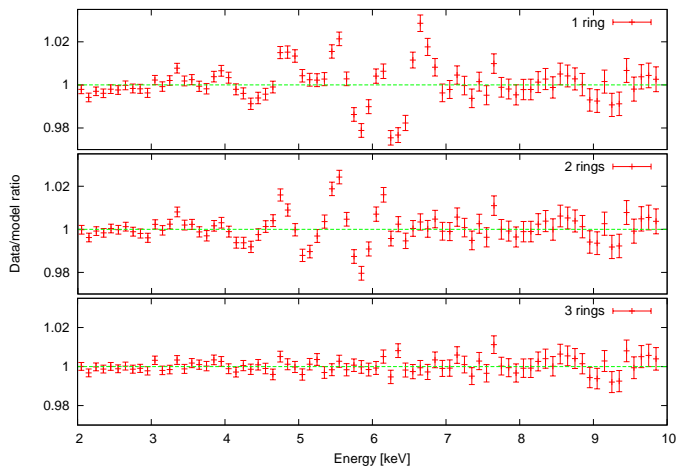}
\caption{Residuals obtained by fitting the fiducial 
spectrum by models with a different assumption about the
number $N$ of accretion rings. Clearly, 
$N=1$ and $N=2$ cases contain some unmodelled features
which allow us to reject these models.
\label{fig:residua}}
\end{center}
\end{figure}

It is interesting to compare our results with a similar scenario for the geometrical and 
kinematical properties of the reflection model by \citet{pariev01}, who also 
based their discussion on the extreme frequency shifts of a spectral line, as 
determined by measured radiation fluxes in the iron line 
wings. In particular, they show contour maps of the extremal 
redshift in two limiting cases -- a non-rotating 
($a=0$, Schwarzschild) and the maximally co-rotating ($a=M$, Kerr) black hole. 
Our graphs are given while keeping the view angle $i$ fixed, 
showing the expected energy of spectral-line wings as the black
hole spin is varied (even to negative values).

We note that there is some unavoidable degeneracy among the model parameters.
However, we will show that this degeneracy can be avoided in situations where
the accretion disc emission is dominated by contributions from a small number
of narrow rings located at well-defined radii. Because the gravitational 
redshift becomes increasingly important as the spin increases and 
the radius of the ring goes to $r_{\rm{}ISCO}$, the rings could
be potentially revealed as features on the wing of the underlying
relativistically broadened line. 

\subsection{Test case}
The assumed source of reflection spectrum is a set of relatively
narrow accretion rings or belts ($\Delta r \sim 0.5 r_{\rm{}g}$) 
representing the emission excess above the standard accretion disc 
spectral line around a rotating (Kerr) black hole. Thanks to the large 
effective area of the proposed detector (designed to reach
$\simeq12\,\mbox{m}^2$, i.e.\ about two orders of magnitude greater 
than that of XMM-Newton near the iron line rest energy), as well as 
a sufficient energy resolution about $200$--$250$ eV,
the accretion rings should be visible when setting realistic
values of the model parameters in our test spectra. 
For the modelling purposes, as the worst case we assume $300$~eV
energy resolution.

\begin{figure*}
\begin{center}
\includegraphics[angle=-90,width=.48\textwidth]{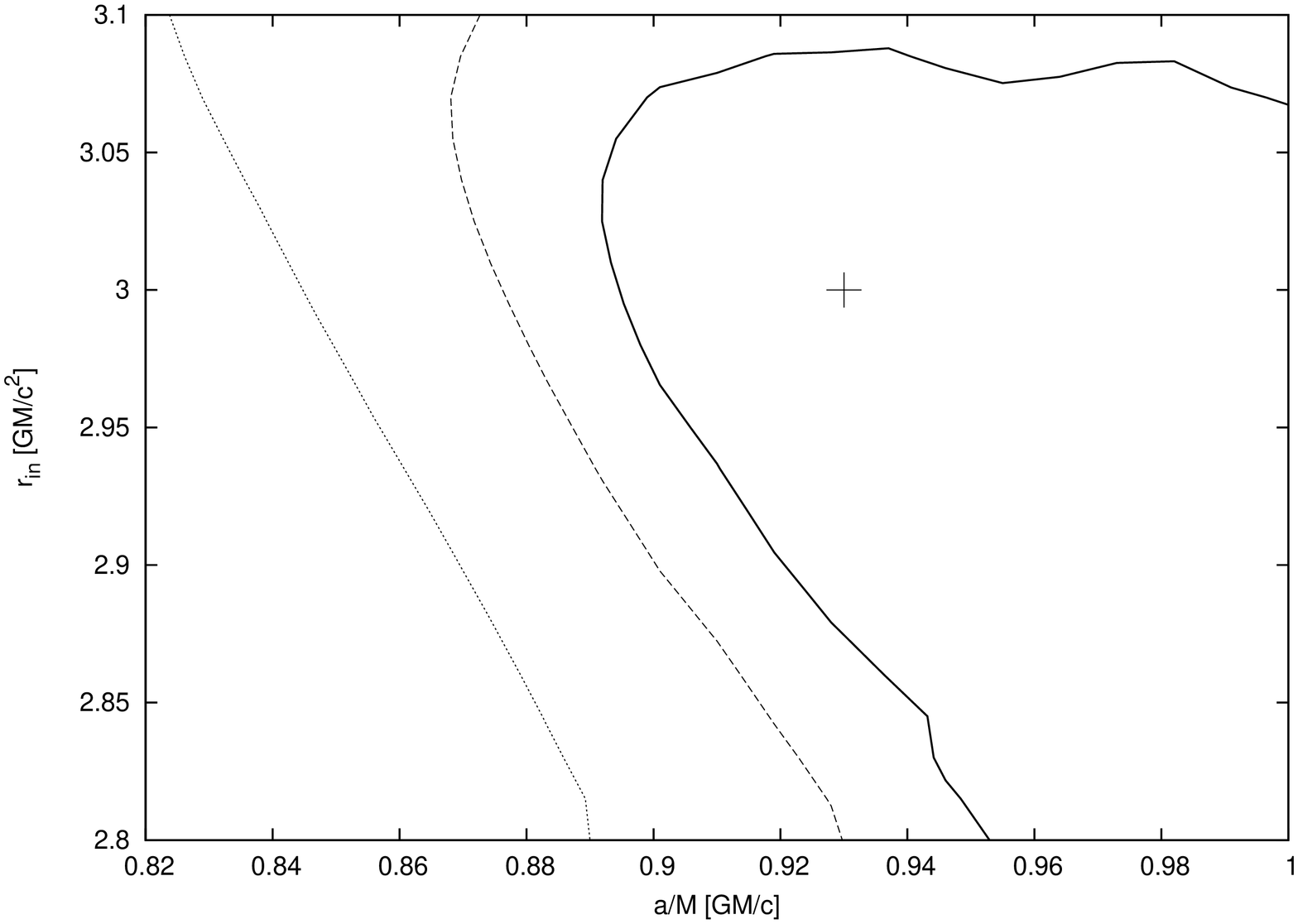}
\hfill
\includegraphics[angle=-90,width=.48\textwidth]{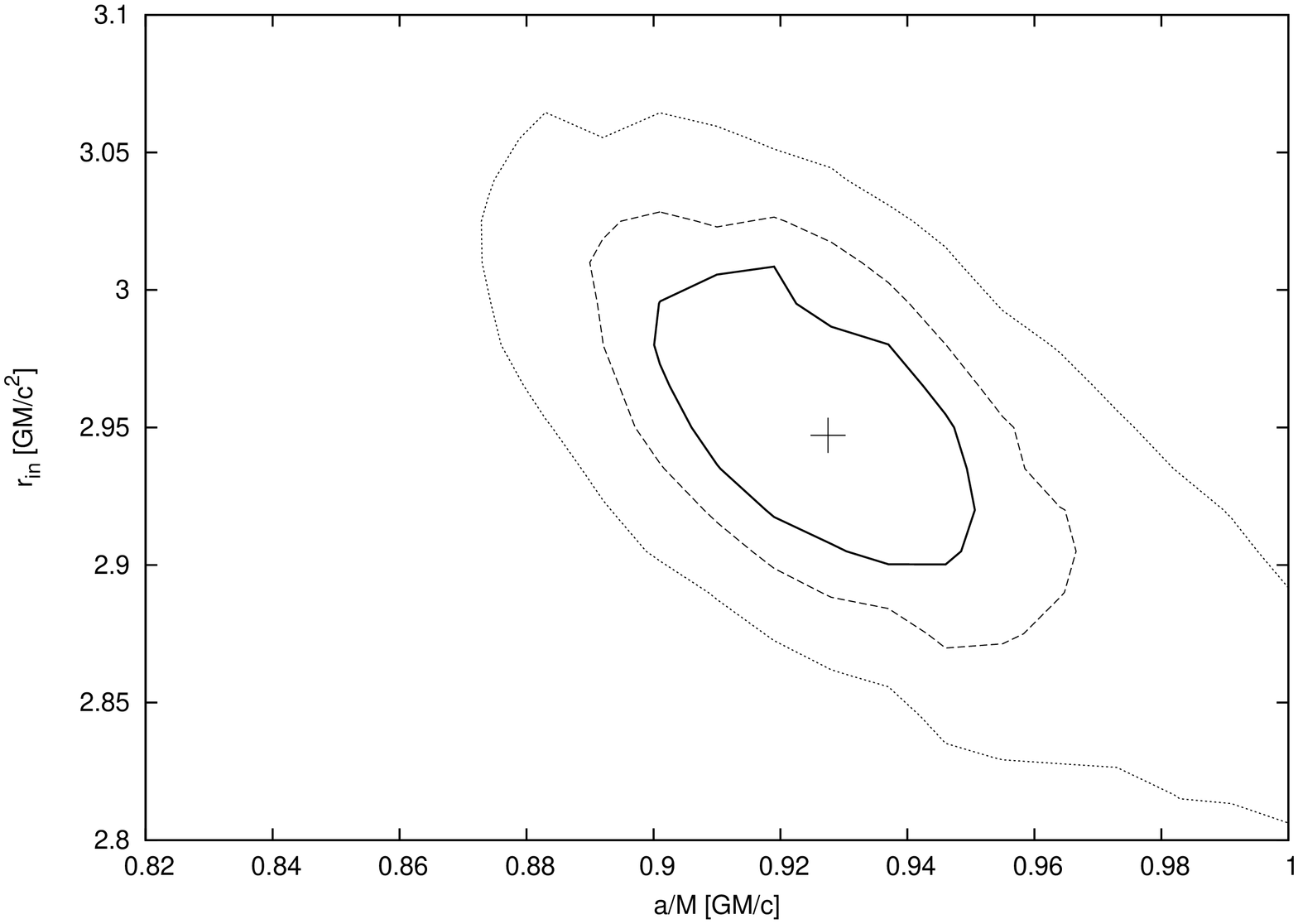}
\caption{Constraints on the best-fit model parameters are derived from the simulation data.
Confidence contours are shown ($1$, $2$, and $3\sigma$) of the inner ring radius $r_{\rm in}$  
vs.\ dimension-less spin $a$. Left panel: 
The best-fit case found using the XMM-Newton response matrix with $100$~ksec exposure time. 
Right panel: The same
analysis performed with the LOFT preliminary response matrix and $20$~ksec exposure.
\label{fig:contours}}
\end{center}
\end{figure*}

In order to demonstrate the feasibility of the mentioned 
scenario we use a preliminary 
response matrix\footnote{See {\tt http://www.isdc.unige.ch/loft/\,.}}
as an example of presumed capability of a future large-collecting-area 
device. This will help us to assess the performance of XMM-Newton versus 
LOFT and our ability of constraining the model.

First, we produced the simulated spectrum by assuming the source flux of 
approximately 1.3 mCrab ($\simeq3\times10^{-11}\,\mbox{erg/cm}^2\mbox{s}$ 
in the energy range $2$--$10$~keV), i.e.\
typical of a nearby bright Seyfert galaxy, such as MCG--6-30-15. 
We assumed a photo-absorbed power law continuum (photon 
index $\Gamma=1.9$, $n_H=4\times10^{21}\mbox{cm}^{-2}$) 
on top of which the
three rings produce a relativistically broadened spectral line (rest
energy $E_{\rm{}rest}=6.4$~keV). 

We assumed the exposure time of
$20$\,ks for the LOFT mission and compared the model spectrum with
the same set-up for $100$\,ks of XMM-Newton spectrum. It is important
to realize that the huge collecting area of LOFT will allow us to constrain
the parameters on a significantly shorter observation time, roughly comparable 
with the orbital time at the corresponding radius. Therefore, the required
exposure does not much exceed reasonable duration of the flares, whereas for
XMM-Newton it had to be significantly longer.

The configuration of three rings is a reasonable (albeit somewhat arbitrary) 
trial assumption,  where
we can test the case of several, but not too many, radially confined excesses
in the disc plane. Also the assumed duration of the observation is somewhat
arbitrarilly set to a value which appears to be reasonably long and realistically
achievable at the same time.

We set the ring widths to be initially $0.5r_{\rm{}g}$, i.e.\ comparable with
the gravitational radius as a typical length-scale of the system. 
Other relevant model parameters were set to: $a=0.93$
(rapidly spinning black hole in prograde rotation), $i=30^{\rm{}o}$
(moderate inclination typical of a Seyfert 1 nucleus) as the initializing
values. Our fiducial model
is therefore {\tt phabs*(powerlaw+4*kyrline)}, i.e.\ a photo-absorbed power-law
continuum and four line components blurred by relativistic effects
(we used XSPEC v.~12.6.0). One of the {\tt kyrline} components
originates over the entire disc surface and it has been fixed to its
defaults parameters ($r_{\rm ISCO}\leq r \leq 400$, radial emissivity index $\alpha=3$);
it provides the main portion of the total line flux (we set about 60\% for definiteness).
The other components represent the emission excesses
from three ring-type regions, each one giving a smaller fraction ($\simeq$10--15\%) 
of the line flux. Naturally, diminishing the line flux produced by the rings
relative to the mean flux of the entire disc and the powerlaw component makes
the determination of $g_{\rm max}$, $g_{\rm min}$ harder and eventually impossible.
We think, however, that the adopted level of the excess emissivity is realistically
possible. Figure~\ref{fig:ringspectrum} shows the data to model ratio
for the fiducial spectrum.
The geometrical location and other parameters of the model components 
will be now submitted to a standard rigorous fitting procedure.

One expects that a single ring
would be easier to recognize in the observed spectrum, while with the
growing number of rings and diminishing separation among them the total
signal should resemble that of a spectral line smeared over the accretion
disc. Can one estimate the model parameters
directly, by locating the energy of the peaks in the spectrum?

\begin{table*}
\begin{center}
\begin{tabular}{cccccclcc}
\multicolumn{1}{c}{\vspace*{-0.9em}Ring} & 
\multicolumn{1}{c}{$g_{\rm{}min}$} & 
\multicolumn{1}{c}{$g_{\rm{}max}$~~~~~} & 
\multicolumn{2}{c}{\rule[-0.8em]{0em}{1.5em} \parbox{3em}{\vspace*{-1em}$r_{\rm in}$}} & 
\multicolumn{1}{c}{~} &
\multicolumn{2}{c}{\parbox{3em}{\vspace*{-1em}$r_{\rm out}$}}   \\  \cline{4-5} \cline{7-8}
\rule[0em]{0em}{1em} & & & $a=0.77$ &$a=1.00$ && $a=0.77$ & $a=1.00$ \\ \hline
1    & 0.36 & 0.81 & 3.1 & 2.8 && 3.7 & 3.4 \\
2    & 0.48 & 0.91 & 4.1 & 3.9 && 4.9 & 4.7 \\
3    & 0.59 & 0.98 & 5.8 & 5.6 && 7.1 & 6.9 \\
\end{tabular}
\caption{Parameters of the model inferred from the energy positions of the
spectral peaks in the test spectrum
from Fig.~\ref{fig:ringspectrum}. We identified the visible features with the horns of
the line components. We imposed the same inclination $i=30$~deg for all three 
rings and required the inferred spin values
to be consistent with each other. The spin turns out to be constrained only partially,
with the values from 0.77 up to 1.00 being consistent with the positions of peaks in the
model spectrum when the radius is set appropriately. The fiducial test spectrum was
generated for rings position at radii $r_{\rm in}=3r_{\rm g}$, $4r_{\rm g}$, and $6r_{\rm g}$, 
respectively. The tabulated
values demonstrate the accuracy of the fitting procedure. See the text for details.}
\label{tab1}
\end{center}
\end{table*}

\begin{figure}
\begin{center}
\includegraphics[angle=-90,width=.49\textwidth]{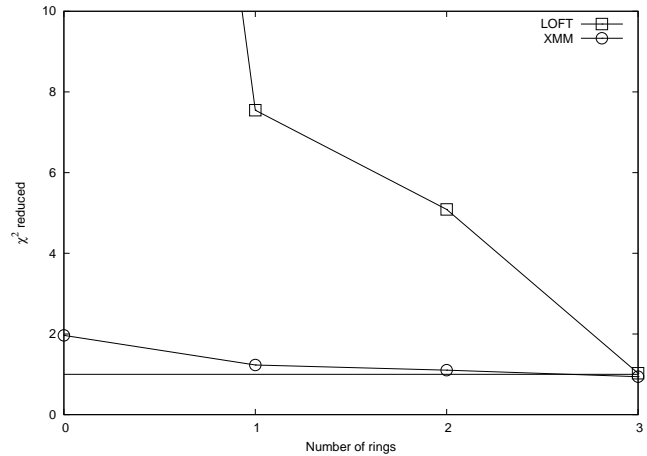}
\caption{Reduced 
$\chi^2$ for different best fits to the fiducial model of 
$N=3$ number of rings. Three cases are shown:
$N=2$ (two rings), $N=1$ (one ring), and $N=0$ 
(the line component produced only by the disc). 
Both $N=1$ and $N=2$ are statistically acceptable 
($\chi^2_{\rm red}\sim1$) with the XMM-Newton 
EPIC-pn response, but they can be excluded
by LOFT. The following parameters were thawed 
for all rings during the fit
procedure: $a$, $i$, $r_{\rm in}$, $r_{\rm{}out}$. The following
parameters were frozen: $E_{\rm rest}$, $\Gamma$, $n_H$. 
\label{fig:chi2}}
\end{center}
\end{figure}

There is a partial degeneracy of the parameter values. In our case 
this exhibits itself by the fact that,
in order to obtain the red peaks of the line in right positions, 
the spin has to be greater than the lower limit of $a=0.77$, 
however, the upper bound remains undetermined (up to $a=1$). 
For $0.77\leq a \leq 1$, i.e.\
up to the maximum spin of the Kerr black hole,
we can reproduce the peaks by rearranging the
ring radii. This is shown in the table by giving two possible values of 
$r_{\rm in}$ and $r_{\rm out}$ that are consistent simultaneously
with the mention minimum
and maximum values of spin. One can see that the uncertainty in the inferred
radii is below 10\%, while for spin the relative error represents about 25\%.

Although the ``visual'' approach to infer the peaks of the line profile
constrains the model quite well, it does not allow us to determine the errors
and to perform the formal statistical confidence contours analysis, so that
apparently accurate values of $g_{\rm max}$ and $g_{\rm min}$ cannot be 
assigned a proper meaning. The spectral fitting should result in a more reliable determination of
the parameters because it employs the whole spectral shape. This appears to be
important especially in the situation when the large effective area allows to resolve the 
broad line. On the other hand,
the spectral fitting tends to be sensitive to the assumed spectral model.
In order to clarify the situation we carry out the following test.

\begin{figure*}
\begin{center}
\includegraphics[angle=-90,width=.49\textwidth]{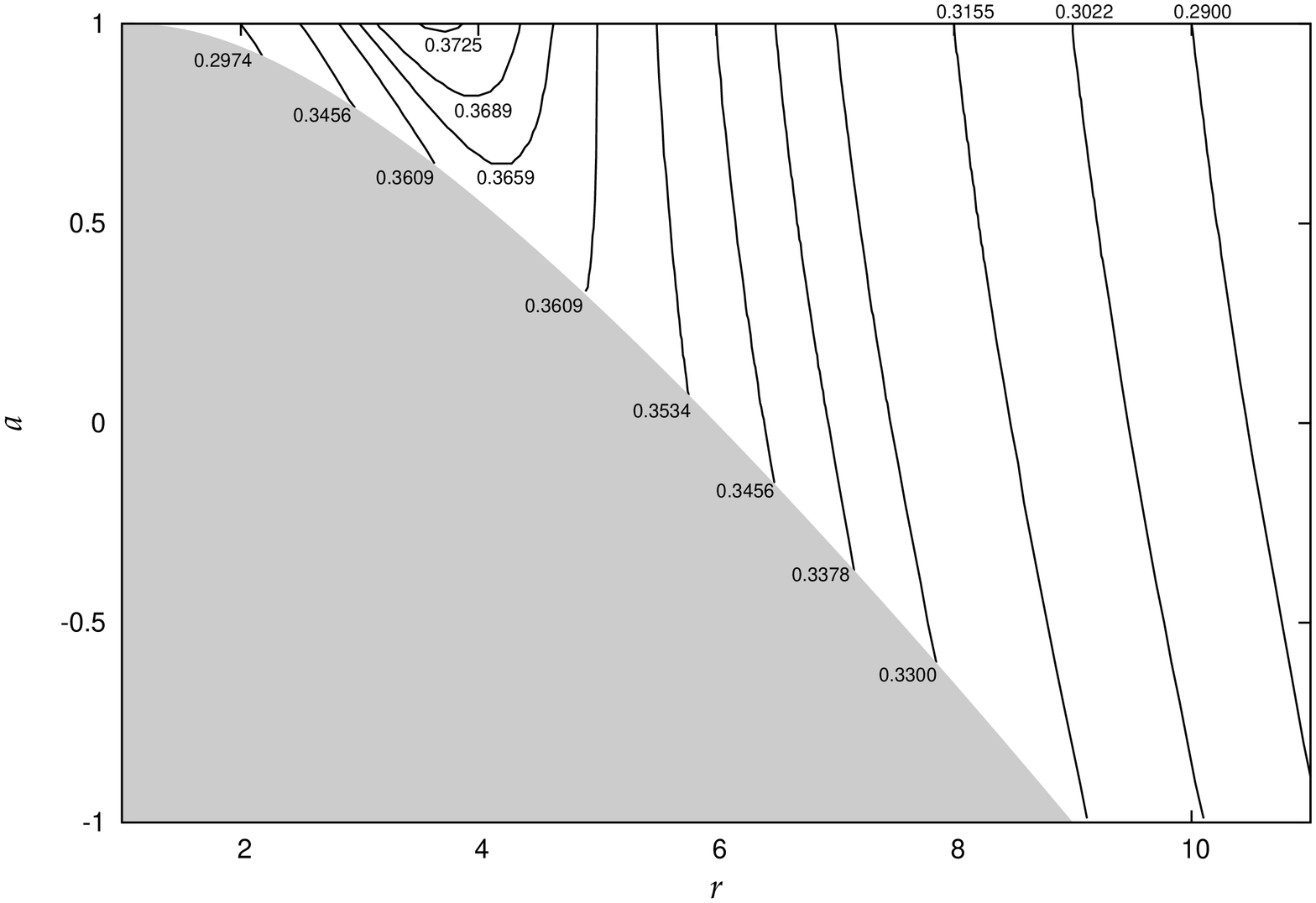}
\includegraphics[angle=-90,width=.49\textwidth]{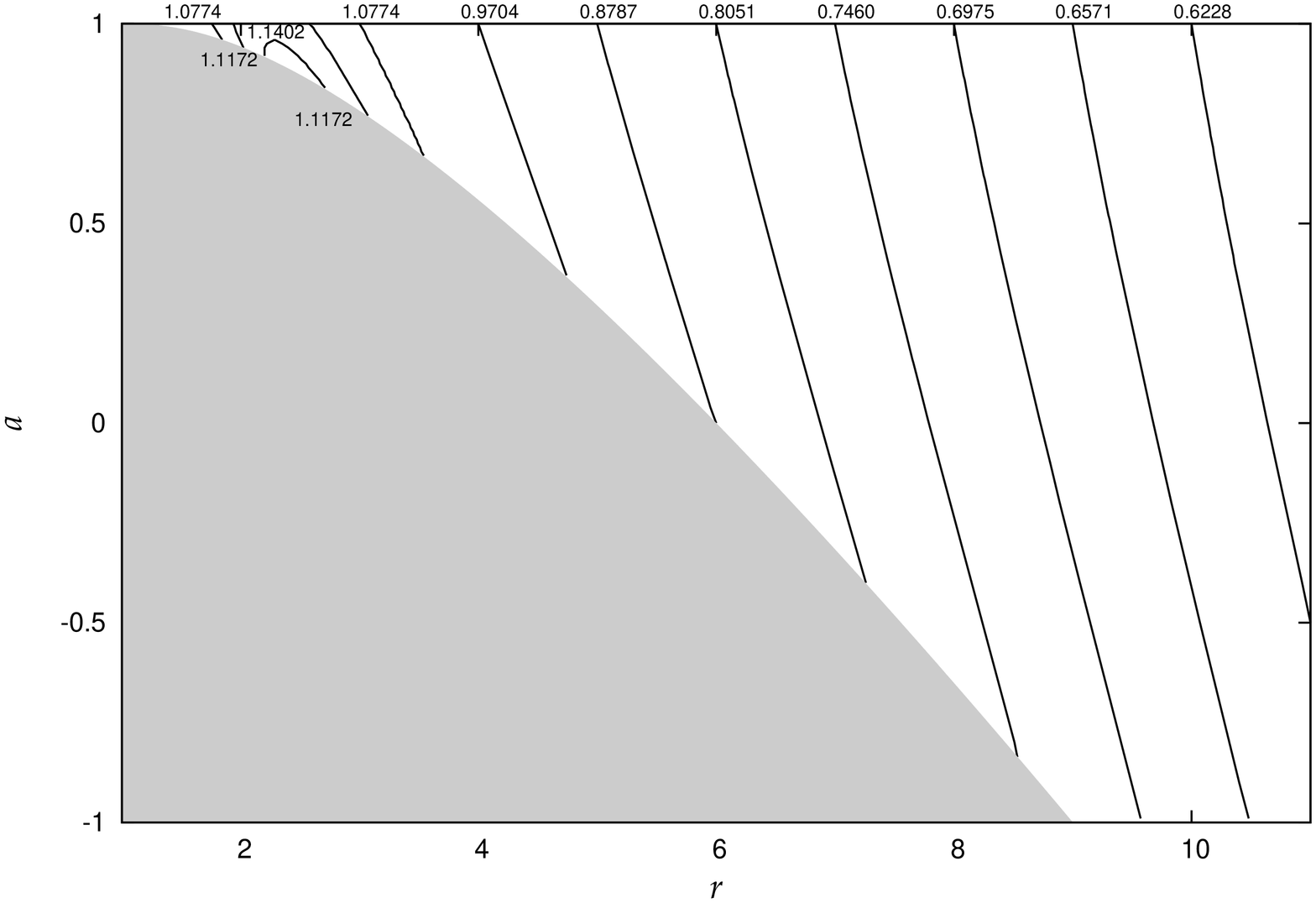}
\caption{Contours of the energy difference $\delta g(r,a;i)$ of the Doppler peaks 
of the observed spectral line profile, as a function
of the dimension-less radius $r$ (units of $GM/c^2$) of the ring and the spin $a$ of
the black hole. $\delta g(r,a;i)$ defines the factor of relativistic broadening of the line. 
Left panel: view angle $i=30$~deg; right panel: $i=75$~deg.
Although the role of spin generally increases as the distance gets smaller,
we notice that $\delta g(r,a;i)$ becomes insensitive to $a$ at a certain distance.
The shaded area represents the region below ISCO.
\label{fig:width}}
\end{center}
\end{figure*}

We subjected the fiducial model to the standard XSPEC fitting
procedure with the aim of recovering the initial model parameters
including their confidence contours. In
particular, we tested if the assumed number of rings can be recovered with a
significant confidence. In figure~\ref{fig:residua}
we show the best-fit residuals with respect to the continuum model, i.e.,
the plot was obtained by
removing the {\tt kyrline} component from the best-fit spectra (we show
unbinned data because any definitive information about the future detector
the quoted energy resolution is still preliminary at present). The
number of rings was changed from the original $N=3$ to 
$N=2$ and $N=1$. As free parameters we re-fitted the spin
$a$, inclination $i$, and the radial width of rings, $\Delta r$. The
rest energy of the line was fixed (it was allowed to vary
in subsequent tests). The position of the rings can be also allowed
to change, although the procedure cannot start too far from correct values
because of local minima in the $\chi^2$ space. 

By setting the number of rings to the fiducial number we
recover correctly the initial values of the parameters as expected, while different numbers
of rings produce worse fits and the parameters converging off the right values
(or not converging at all). 
We examined also the case of the line emission spread over the entire disc
surface (this represents the standard disc-line model), but the resulting fit was bad.

In order to compare the results expected from LOFT with those that
can be reached with currently available data, we carried out the same
procedure as described above also with the XMM-Newton response file of
the EPIC-pn camera. 
Figure \ref{fig:contours} demonstrates the expected accuracy with which 
the model parameters are constrained.
We plot the confidence contours for the inner ring radius versus
the black hole spin. With the 
same exposure and the source brightness for both instruments, 
LOFT will allow us to set
much tighter constraints on the best-fit values.


In figure~\ref{fig:chi2} we show the resulting
best-fit $\chi^2_{\rm{}red}$ values. Decreasing the number of rings from
the right value $N=3$ obviously degrades goodness of the fit. 
For XMM-Newton the
best-fit $\chi^2_{\rm{}red}$ values are lower, meaning that the model
parameters are constrained at deteriorated confidence compared to
LOFT. Both $N=1$ and $N=2$ tests give (wrongly) acceptable 
fits for XMM-Newton, while they can be clearly rejected with LOFT. 
Naturally, $N=3$ gives a good fit because our artificial data were
created with this value, so we only reproduce the original input.
Increasing the number of rings above $N=3$
produces formally acceptable statistics, but the number of model
parameters then exceeds the necessary minimum. This indicates an encouraging
improvement that could be achieved for this particular problem with the
improved sensitivity of LOFT.

\section{Discussion}
\label{sec:Discussion}
Parameter constraints could be improved also in our test 
case of XMM-Newton, if the observation duration can be made longer and the number
of collected photons correspondingly larger. Because in practice one cannot
prolong the duration above some reasonable limit, the increase of
the collecting area of the LOFT detector is the most important aspect which
improves the parameter constraints. The energy resolution 
of the detector is a less critical factor because the effect of relativistic broadening 
causes the line width to be much broader than this limit by a significant margin.

We remark that the special choice of $N=3$ was needed for illustration
purposes described above, but it does not mean any crucial limitation that would be
essential for the described idea or the method. Naturally, a lower number of rings 
is easier to be recognized in the spectrum and its parameters to be
reconstructed correctly. On the other hand, in the case of large $N$ the
magnitude of the resulting peaks above the baseline profile becomes
relatively small and the situation turns into the problem discussed by
\citet{wilkins11}, i.e.\ the disc-line emission imagined as consisting of
infinitesimal rings extending all the way from the inner to the outer
rim of the disc.

We asked ourselves whether the energy position of the observed peaks 
determines the system parameters $r$, $i$, and $a$ in any unique
way, or if instead the proposed procedure of reconstructing the parameter values
converges to several different results, depending on the initial guess
of the parameter values. In fact, the latter is true. 
Therefore we need to set addition constraints
in order to help XSPEC to find the right answer.
We can demonstrate this complication in another way,
by plotting the observed energy difference 
$\delta g(r,a;i)$ of the line peaks in figure~\ref{fig:width}: $\delta g\equiv 
g_{\rm max}-g_{\rm min}$. Notice that the dependence of $\delta g$ on spin
is ambiguous in some parts of the parameter space.

In some situations -- namely, for small radii and large spins -- the 
energy separation of the two peaks the observed profile
is not in one-to-one correspondence to model parameters; 
it is instead a double-valued quantity with respect to $a$ (as can be seen
in the top left corner of the fig.~\ref{fig:width} left panel). 
Naturally, at large radii and small inclinations 
the spin dependence on $\delta g$ is weak, and so the practical
use of the method becomes compromised. On the other hand, we do see 
the range of the parameter space (typically for radii between $r\simeq 4$ and 
$\simeq 12$), where the spin dependence is non-negligible and
the possibility of reconstructing the parameters appears quite promising. 
We can also notice an
interesting conspiracy when the contour-line takes almost straight
vertical direction (e.g.\ in the same plot for $i\simeq30$~deg, $r\simeq~5GM/c^2$),
implying that spin is degenerate with respect to 
radius for that given inclination.

Also, a narrow Fe line -- so ubiquitous in the X-ray spectra of type-1 
AGN -- will be unresolved by LOFT due to the poor energy resolution
of its Large Area Detector (LAD). This could lead to confusion with the blue wing of the relativistic 
line if it happens to come close to the rest energy of the neutral line. 
However, it does not mean an unsurmountable complication because these narrow 
features are thought to originate via scattering by a distant reflector (e.g.\ a torus of matter
located at several thousand gravitational radii, or more). Therefore, the narrow
component does not vary on short time-scales and its influence can be taken
into account by estimating the flux from other, high-energy resolution spectroscopy,
and including it in the model. Indeed, most of candidate objects for a relativistic line were 
already observed by XMM-Newton \citep{nandra07,delacalle10}, although a concurrent 
detection by LOFT together with another high-energy resolution mission
would be the best strategy.

Naturally, the problem of confusion or non-uniqueness would appear
if the broad line is produced farther out from ISCO (several tens of $r_{\rm g}$
and more), so that the observed width is only moderate and comparable with the 
detector intrinsic energy resolution; for such cases the LOFT will not be suitable. 
We remark that the aforementioned simulations were performed
for the response matrix resolution of 300~eV. It is currently expected that a better
resolution of about $200$--$250$~eV can be realistically achieved \citep{feroci11}, and so the obtained
constraints should come out as even more favourable compared to the results 
presented here. As an example, we constructed the confidence contours of 
Fig.~\ref{fig:contours}, but with the energy resolution of $180$~eV (an
optimistic expectation), and we checked that the contour levels come out very 
similar to those shown for $300$~eV resolution in the right panel. Both sets
of best-fit values are consistent with each other.

The above-given simulations suggest that the mentioned
features should be detectable around the iron line 
rest energy in bright accreting supermassive black 
holes. The foreseen energy resolution of the 
detector is sufficient to reveal moderately broad excesses emerging 
on top of the underlying continuum (unless the resolution becomes significantly 
worse than the currently discussed value about 200 eV). The signal-to-noise 
level would be more of a concern especially if the background level were not 
determined with sufficient confidence. Our example demonstrates that 
bright AGNs \citep[such as MCG--6-30-15 during a prominent 
flare of][]{ponti04} should be accessible to this kind 
of study. To this end, the background model is required to be stable 
to approximately $5$\% accuracy.\footnote{We tested the expected
impact of background contamination by applying
{\tt corrnorm} to the background file in XSPEC. Randomizing the background
at 3, 5, and 10\% levels, respectively, we checked that the parameters
of our fiducial model are well-reproduced in the first two cases, while the 10\%
inaccuracy of the background would seriously degrade the best fit constraints.}

The expected signal from AGNs will be background dominated with 
LOFT. Therefore, for the applications discussed in this paper, the precise 
determination of the background model is important, 
especially near the iron line rest energy. More detailed studies
of the background impact are needed, however, currently available 
estimations \citep{feroci11} suggest sufficient stability of the background
model on the time-scales exceeding the orbital periods in
the inner regions of AGN accretion discs.
Promising targets can be picked up from the 
list by \citet[][and further references cited therein]{demarco09}; these 
objects have already exhibited the presence of transient narrow-line features
that could be relevant for our study.

Naturally, Galactic black holes in X-ray binaries represent another category 
of potential targets, assuming one can identify a suitable source with the flare-induced 
reflection signatures, similar to those seen in AGNs. According to the current 
specifications, the temporal resolution does 
not pose serious limitations (thanks to the large 
collecting area and sufficiently fast response time), even though the 
characteristic time-scale decreases proportionally to the mass, i.e., by 6--7 orders of 
magnitude shorter for Galactic black holes compared to supermassive black holes. 
On the other hand, high ionisation levels are typical of Galactic black 
holes, and these may prevent successful detections despite the favorable source 
brightness that exceeds typical AGNs.

The adopted scenario has clearly an interesting potential to discriminate
between the reflection model of spectral features from the case when the
obscuration by intervening material and the resulting absorption are the main agents 
modulating the power-law continuum and shaping the observed spectrum.

In summary, we arrive at the conclusion that the fiducial
values of the model parameters can be reconstructed with the anticipated
capabilities of LOFT. Although in this paper we considered the spectral features formed by
azimuth-integrated rings, our preferred scenario of forming these rings is by
orbital motion of spatially localized flares. We note that the LOFT mission is
dedicated primarily to timing studies. Indeed, it should provide us with
light curves of flares to much better time resolution than any of its 
predecessors. Temporal profiles of the flare onset and the subsequent decay
exhibit some signatures of model parameters \citep[see fig.\ 7 in][]{goosmann07}, 
but these are beyond current observational capabilities.
Once resolved, they will give us a complementary information that can be 
compared against spectral studies discussed in the paper.

\section{Conclusions}
We explored a possibility of studying relatively indistinct excesses 
on top of  the relativistically broadened spectral line profile. Unlike the 
main body of the broad line, thought to originate from a whole
extended region of the accretion disc, we modelled these features 
as the emission-line components arising in well-confined
radial distance in the accretion disc. We also suggested that
some energetically narrow features could be explained as a signature of
spatially localized irradiation by magnetic reconnection flares above 
the underlying accretion disc. Because of prevailing rotational motion of
the accreted material and unavoidably prolonged duration
of observations necessary to collect enough photons, even such
localized events should reveal themselves in the observed spectrum 
as emission rings.

We found that the presence of about 10--15\% excess emission
of the line flux originating from a ring of a moderate width (typically a fraction of 
gravitational radius) close to the black hole could be tested by the proposed
LOFT satellite. The effective area of this mission is designed to be large enough to
allow the model parameters to be constrained despite its limited
energy resolution. This will significantly reduce
the degeneracy of the model parameters -- a notorious obstacle complicating
interpretations of the current data. One needs to note that the ability of converging
to the correct parameter values could be compromised if the background level is significantly
increased above our default estimation, so this issue will need further 
investigation. In fact, bright Galactic stellar-mass black-hole candidates may be more
suitable sources, as AGN will be always dominated by the background.

In this paper 
we concentrated ourselves to the supermassive black hole especially because in this case the
base-line model can be assumed in a relatively simple form and the iron-line complex has
been commonly detected in AGN, nevertheless, bright Galactic
binaries with accurately modelled spectral components may be eventually a more
suitable category of targets. At this stage we could only verify that the anticipated
background level does not pose a serious limitation for the example described above,
assuming that the background model is itself known with the sufficient accuracy. This
issue will need further investigation especially because the background level is 
expected to vary with energy, and so, in principle, inaccuracy of the background model
could arise.

A typical double-horn profile gives us an opportunity to
determine the parameters by measuring the energy shifts of the 
features within the broad spectral line wings. The observed energy of these 
features is well-defined (assuming a sufficiently narrow radial extent of 
the rings). Naturally, the inferred parameters from different rings
should be consistent with each other
(e.g.\ the same inclination, assuming that the rings reside in the same 
equatorial plane). This will help us to distinguish the disc--line
geometry of the line emitting region from alternative options,
such as the lines originating in outflows and jets \citep{wang00},
spiral waves \citep{hartnoll02},
or geometrically thick non-Keplerian tori \citep{fuerst07}.

Even in situations when the contribution of rings is only moderate
and the energy shifts cannot be determined immediately from the
secondary peaks merely by inspecting the spectrum, the fitting procedure
can be employed to reconstruct the model parameters. This requires employing a 
physically substantiated model of the spectrum and using the entire
profile of the broad line and the continuum. Naturally, in the latter case
the assumptions about the intrinsic spectrum are essential. 

The concept seems to be well suited to study the effects
which are often coined as ``strong gravity'' in the astrophysical context. Despite
a limited energy resolution and various convergence issues of the spectral 
fitting procedure discussed in the paper above, the large effective area offers a 
significant progress over what can be presently achieved with XMM-Newton 
not only in the case of  timing studies but also for the spectroscopy of truly broad
relativistic spectral features.

\section*{Acknowledgments}
We thank Dr.\ Alessandra de Rosa for advice concerning the background estimation,
and an unknown referee for useful comments.
This research is supported by the ESA PECS project No.~98040. VK and MD
gratefully acknowledge support from the Czech Science Foundation grant
205/07/0052 and the Ministry of Education project ME09036. VS
is supported by the doctoral grant GA\v{C}R No.\ 205/09/H033.
JS acknowledges support from the grant GA\v{C}R 202/09/0772.
Part of this work has been carried out within the COST Action MP0905 -- 
Black Holes in a Violent Universe.

{}

\label{lastpage}
\end{document}